\documentclass[runningheads]{llncs}
\usepackage{graphicx}
\usepackage{hyperref}

\usepackage{enumitem}
\usepackage{siunitx}

\usepackage{amsmath}
\usepackage{amssymb}
\DeclareMathOperator*{\argmax}{arg\,max}

\usepackage{color}

\usepackage[acronym,shortcuts]{glossaries}
\newacronym{MEA}{MEA}{microelectrode array}
\newacronym{NFB}{NFB}{nerve fiber bundle}
\newacronym{RGC}{RGC}{retinal ganglion cell}
\newacronym{RP}{RP}{retinitis pigmentosa}
\glsdisablehyper  

\begin{document}
\title{Greedy Optimization of Electrode Arrangement for Epiretinal Prostheses}
\author{Ashley Bruce\inst{1} \and
Michael Beyeler\inst{1,2}\orcidID{0000-0001-5233-844X}}
\index{Bruce, Ashley}
\index{Beyeler, Michael} 
\authorrunning{Bruce \& Beyeler}
\institute{Department of Computer Science \and Department of Psychological \& Brain Sciences \\
University of California, Santa Barbara, CA 93106
\email{\{ashleybruce,mbeyeler\}@ucsb.edu}}
\maketitle              
\begin{abstract}
Visual neuroprostheses are the only FDA-approved technology for the treatment of retinal degenerative blindness. 
Although recent work has demonstrated a systematic relationship between electrode location and the shape of the elicited visual percept, this knowledge has yet to be incorporated into retinal prosthesis design, where electrodes are typically arranged on either a rectangular or hexagonal grid.
Here we optimize the intraocular placement of epiretinal electrodes using dictionary learning.
Importantly, the optimization process is informed by a previously established and psychophysically validated model of simulated prosthetic vision.
We systematically evaluate three different electrode placement strategies across a wide range of possible phosphene shapes and recommend electrode arrangements that maximize visual subfield coverage.
In the near future, our work may guide the prototyping of next-generation neuroprostheses.

\end{abstract}
\section{Introduction}

Current visual neuroprostheses consist of a \ac{MEA} implanted into the eye or brain that is used to electrically stimulate surviving cells in the visual system in an effort to elicit visual percepts (``phosphenes'').
Current epiretinal implant users perceive highly distorted percepts, which vary in shape not just across subjects, but also across electrodes \cite{erickson-davis_what_2021,luo_long-term_2016}, and may be caused by incidental stimulation of passing \acp{NFB} in the retina \cite{beyeler_model_2019,rizzo_perceptual_2003}.
However, this knowledge has yet to be incorporated into prosthesis design.

To address this challenge, we make the following contributions:

\begin{enumerate}[topsep=0pt,itemsep=-1ex,partopsep=0pt,parsep=1ex,leftmargin=36pt]
    \item We optimize the intraocular placement of epiretinal electrodes using dictionary learning. Importantly, this optimization process is informed by a previously established and psychophysically validated model of simulated prosthetic vision \cite{beyeler_pulse2percept_2017,beyeler_model_2019}.
    \item We systematically evaluate three different electrode placement strategies across a wide range of possible phosphene shapes and recommend electrode arrangements that maximize visual subfield coverage.
\end{enumerate}

\section{Related Work}

\begin{figure}[b!]
\centering
\includegraphics[width=0.9\columnwidth]{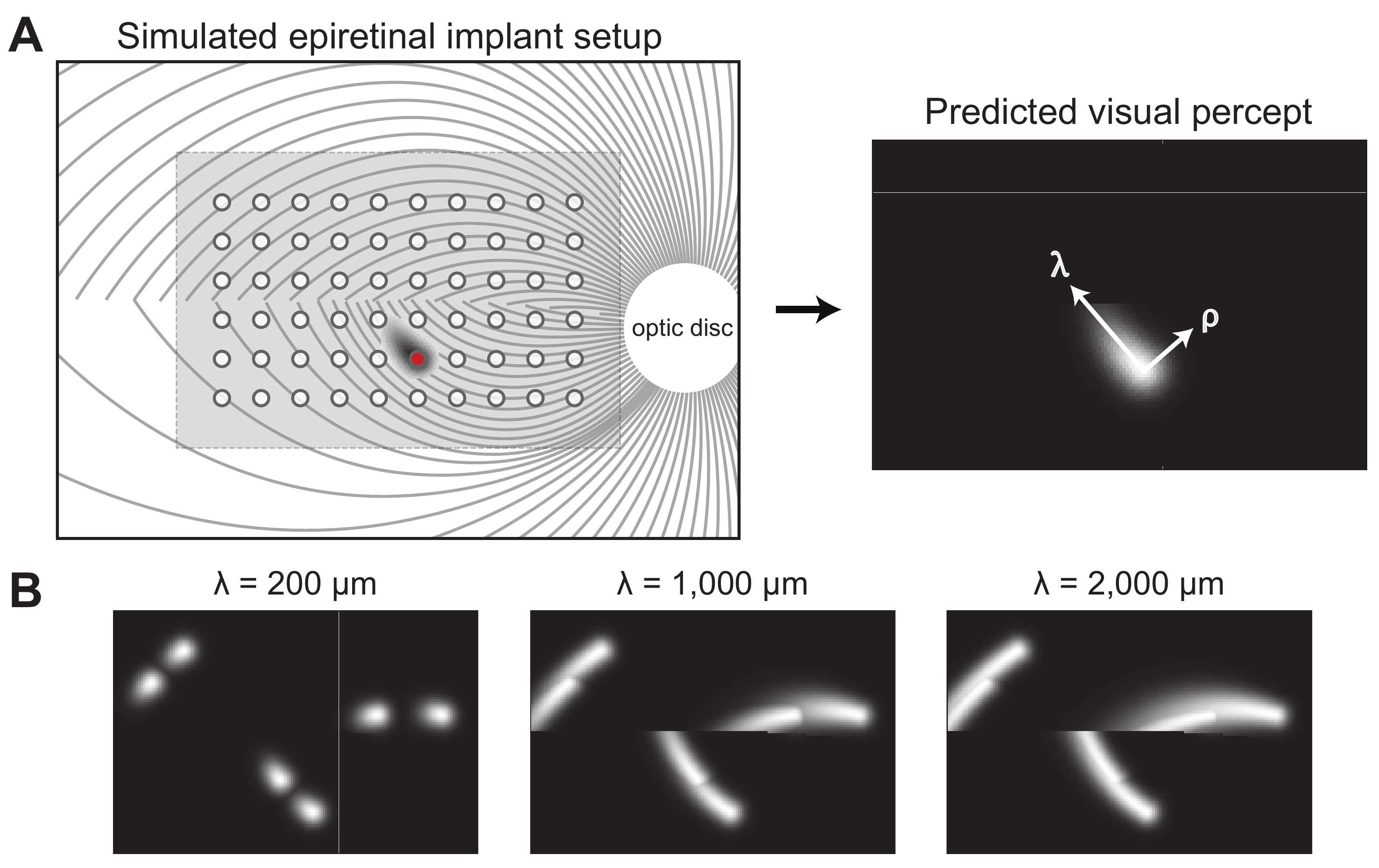}
\caption{\textbf{A)} Axon map model. 
    \emph{Left}: Electrical stimulation (red disc) of a \acs{NFB} (gray lines) leads to tissue activation (dark-gray shaded region) elongated
    along the \acs{NFB} trajectory away from the optic disc (white circle).
    The light-gray shaded region indicates the visual subfield that is being simulated.
    \emph{Right}: The resulting visual percept appears elongated as well; its shape can be described by two parameters, $\lambda$ (spatial extent along the \acs{NFB} trajectory) and $\rho$ (spatial extent perpendicular to the \acs{NFB}).
    \textbf{B)} 
    As $\lambda$ increases, percepts become more elongated and start to overlap.
}
\label{fig:axonmap}
\end{figure}

Sensory neuroprostheses such as retinal and cochlear implants are emerging as a promising technology to restore lost sensory function.
These devices bypass the natural sensory transduction mechanism and provide direct electrical stimulation of (retinal or auditory) nerve fibers that the brain interprets as a (visual or auditory) percept.
However, perceptual distortions can occur when multiple electrodes stimulate the same neural pathways \cite{beyeler_learning_2017,wilke_electric_2011}, and electrode placement has been shown to impact the quality of visual and hearing outcomes \cite{beyeler_model_2019,finley_role_2008}.

However, this information has yet to be incorporated into neuroprosthetic device design and surgical placement, with most patients having less-than-optimal \ac{MEA} placement \cite{beyeler_model-based_2019,chakravorti_further_2019}.
Whereas previous studies have optimized the shape of individual electrodes \cite{rattay_effective_2004}, which electrodes to activate in order to produce a desired visual response \cite{bratu_graph-based_2020,granley_hybrid_2022,shah_optimization_2019}, or the overall implant placement \cite{beyeler_model-based_2019}, we are unaware of any studies that have attempted to optimize the placement of individual electrodes within an implant.

In the case of retinal implants, recent work has demonstrated that phosphene shape strongly depends on the retinal location of the stimulating electrode \cite{beyeler_model_2019}.
Because \acp{RGC} send their axons on highly stereotyped pathways to the optic nerve, an electrode that stimulates nearby axonal fibers would be expected to antidromically activate \ac{RGC} bodies located peripheral to the point of stimulation, leading to percepts that appear elongated in the direction of the underlying \ac{NFB} trajectory (Fig.~\ref{fig:axonmap}A, \emph{right}).
Using a simulated map of \acp{NFB}, Reference \cite{beyeler_model_2019} was thus able to accurately predict phosphene shape for various users of the Argus Retinal Prosthesis System (Second Sight Medical Products, Inc.), by assuming that an axon's sensitivity to electrical stimulation:
\begin{enumerate}[topsep=0pt,itemsep=-1ex,partopsep=0pt,parsep=1ex,leftmargin=16pt,label=\roman*.]
    \item decays exponentially with decay constant $\rho$ as a function of distance from the stimulation site,
    \item decays exponentially with decay constant $\lambda$ as a function of distance from the cell body, measured as axon path length.
\end{enumerate}

As can be seen in Fig.~\ref{fig:axonmap}B, electrodes near the horizontal meridian are predicted to elicit circular percepts, while other electrodes are predicted to produce elongated percepts that will differ in angle based on whether they fall above or below the horizontal meridian.
In addition, the values of $\rho$ and $\lambda$ dictate the size and elongation of elicited phosphenes, respectively, which may drastically affect visual outcomes.
Specifically, if two electrodes happen to activate the same \ac{NFB}, they might not generate two distinct phosphenes (Fig.~\ref{fig:axonmap}B, \emph{right}).

Instead of arranging electrodes such that they efficiently tile the \emph{retinal surface}, a better approach might thus be to arrange electrodes such that the \emph{elicited percepts} effectively tile the \emph{visual field}.
In the following, we will demonstrate that this is equivalent to finding the smallest set of electrodes that cover a desired portion of the visual field (here termed a visual subfield).
However, this strategy can be readily applied wherever there is a topological mapping from stimulus space to perceptual space (\emph{e.g.}, visual, auditory, tactile stimulation).
Incorporating this knowledge into implant design could therefore be indispensable to the success of future visual neuroprostheses.

\section{Methods}

\subsection{Phosphene Model}

Let $\mathcal{E} = \{e_1, \ldots, e_N\}$ be the set of $N$ electrodes in a \ac{MEA}, where the $i$-th electrode $e_i = (x_i, y_i, r_i)$ is described by its location on the retinal surface $(x_i, y_i)$ and its radius $r_i > 0$.
For example, Argus II can be described by $\mathcal{E}_\mathrm{Argus II}$, where $N = |\mathcal{E_\mathrm{Argus II}}| = 60$, $r_i = \SI{122.5}{\micro\meter}$ $\forall i$, and $(x_i, y_i)$ are spaced \SI{575}{\micro\meter} apart on a rectangular grid.
We do not assume any particular ordering of $\mathcal{E}$.

Furthermore, let $\mathcal{S} = \{(s_1, \ldots, s_k)\}$ be the set of stimuli sent to $k \le N$ electrodes in the \ac{MEA}, where the $i$-th stimulus $s_i = (e_i, a_i)$ is described by an electrode $e_i \in \mathcal{E}$ and its corresponding activation function $a_i$.
In practice, $a_i$ may be a biphasic pulse train of a given duration, pulse amplitude, and pulse frequency.
However, for the purpose of this study, we limited ourselves to spatial activation values $a_i \in \mathbb{R}_{\ge 0}$, which did not contain a temporal component.

A phosphene model $\mathcal{M}$ then takes a set of stimuli $\mathcal{S}$ as input and outputs a visual percept $p \in \mathbb{R}^{H \times W}_{\ge 0}$, which is a height ($H$) $\times$ width ($W$) grayscale image.
In general, $\mathcal{M}$ is a nonlinear function of $\mathcal{S}$ and depends on subject-specific parameters $\theta_\mathrm{subject}$ such as $\rho$ and $\lambda$; thus $p = \mathcal{M}(\mathcal{S}; \theta_\mathrm{subject})$.

For the purpose of this study, we used \emph{pulse2percept} 0.8.0, a Python-based simulation framework for bionic vision \cite{beyeler_pulse2percept_2017} that provides an open-source implementation of the axon map model \cite{beyeler_model_2019}, as described in the previous section.
Constrained by electrophysiological and psychophysical data, this model predicts what a bionic eye user should ``see'' for any given set of stimuli $\mathcal{S}$.

\subsection{Dictionary Selection}

\subsubsection{Problem Formulation}

Let $\mathcal{T}$ now be the set of all nonoverlapping epiretinal electrodes, and $\mathcal{D}$ be a subset of those; that is, $\mathcal{D} \subset \mathcal{T}$, where $|\mathcal{D}| \ll |\mathcal{T}|$.
In the dictionary selection problem, we are interested in finding the dictionary $\mathcal{D}^*$ that maximizes a utility function $F$ (see next subsection):
\begin{equation}
    \mathcal{D^*} = \argmax_{|\mathcal{D}| \le k} F(\mathcal{D}),
\end{equation}
where $k$ is a constraint on the number of electrodes that the dictionary can be composed of.

This optimization problem presents combinatorial challenges, as we have to find the set $\mathcal{D}^*$ out of exponentially many options in $\mathcal{T}$.
However, we will only consider utility functions $F$ with the following properties:
\begin{enumerate}[topsep=0pt,itemsep=-1ex,partopsep=0pt,parsep=1ex,leftmargin=18pt,label=\roman*.]
    \item The empty set has zero utility; that is, $F(\emptyset) = 0$.
    \item $F$ increases monotonically; that is, whenever $\mathcal{D} \subseteq \mathcal{D'}$, then $F(\mathcal{D}) \le F(\mathcal{D'})$.
    \item $F$ is approximately submodular; that is, there exists an $\varepsilon$ such that whenever $\mathcal{D} \subseteq \mathcal{D'} \subseteq \mathcal{T}$ and an electrode $e \in \mathcal{T} \setminus \mathcal{D'}$, it holds that $F(\mathcal{D} \cup \{e\}) - F(\mathcal{D}) \ge F(\mathcal{D'} \cup \{e\}) - F(\mathcal{D'}) - \varepsilon$.
    This property implies that adding a new electrode $e$ to a larger dictionary $\mathcal{D'}$ helps at most $\varepsilon$ more than adding $e$ to a subset $\mathcal{D} \subseteq \mathcal{D'}$.
\end{enumerate}
For utility functions with the above properties, Nemhauser et al.~\cite{nemhauser_analysis_1978} proved that a simple greedy algorithm that starts with the empty set $\mathcal{D}_0 = \emptyset$, and at every iteration $i$
 adds the element
 \begin{equation}
     d_i = \argmax_{d \in \mathcal{T} \setminus \mathcal{D}} F(\mathcal{D}_{i-1} \cup \{d\}),
     \label{eq:dict-baseline}
 \end{equation}
where $\mathcal{D}_i = \{d_1, \ldots, d_i\}$, is able to obtain a near-optimal solution.

\subsubsection{Utility Function}

Ideally, the utility function $F$ would directly assess the quality of the generated artificial vision.
As a first step towards such a quality measure, we considered the ability of a set of electrodes to lead to phosphenes that cover a specific visual subfield (i.e., the gray shaded region in Fig.~\ref{fig:axonmap}).
We would thus activate every electrode in $\mathcal{D}$, represented by $\mathcal{S}_\mathcal{D}$, and calculate the percept $p_\mathcal{D} = \mathcal{M}(\mathcal{S}_\mathcal{D}; \theta_\mathrm{subject})$.
Then $F$ was given as the visual subfield coverage; that is:
\begin{equation}
    F(\mathcal{D}) = \sum_{w=1}^{w=W} \sum_{h=1}^{h=H} p_{hw} \ge \varepsilon,
\end{equation}
where $W$ and $H$ were the width and height of the percept, respectively, $\varepsilon = 0.1 \max_{\forall h, w}(p_{hw}$), and $F \in [0, HW]$.

\subsubsection{Dictionary Selection Strategies}
To find $\mathcal{D}^*$, we considered three different strategies:

\begin{itemize}[topsep=0pt]
    \item {\bf No Overlap:} At each iteration $i$, $d_i$ was chosen according to Eq.~\ref{eq:dict-baseline}.
    \item {\bf Not Too Close:} To consider manufacturing constraints, we modified Eq.~\ref{eq:dict-baseline} above to enforce that all electrodes were placed at least $c= \SI{112.5}{\micro\meter}$ apart from each other:
    \begin{equation}
        d_i = \argmax_{d \in \mathcal{T} \setminus \mathcal{D} \; \mathrm{s.t.} \; ||d, d_j||_2 \ge c \; \forall d_j \: \in \: \mathcal{D} \setminus d} F(\mathcal{D}_{i-1} \cup \{d\}).
        \label{eq:dict-no-overlap}
    \end{equation}
    \item {\bf Max Pairwise Distance:} As electrical crosstalk is one of the main causes of impaired spatial resolution in retinal implants \cite{wilke_electric_2011}, we further modified Eq.~\ref{eq:dict-no-overlap} to place electrodes as far away from each other as possible:
    \begin{equation}
        d_i = \argmax_{d \in \mathcal{T} \setminus \mathcal{D} \; \mathrm{s.t.} \; ||d, d_j||_2 \ge c \; \forall d_j \: \in \: \mathcal{D} \setminus d} F(\mathcal{D}_{i-1} \cup \{d\}) + \alpha \sum_{d_i \in \mathcal{D} \setminus d} ||d, d_i||_2,
        \label{eq:dict-max-dist}
    \end{equation}
    where $\alpha = \num{1e-4}$ was a scaling factor. 
\end{itemize}

\subsubsection{Implementation Details}

For the sake of feasibility, we limited $\mathcal{T}$ to electrodes placed on a finely spaced search grid, $x_i \in [-3000, 3000] \: \SI{}{\micro\meter}$ and $y_i \in [-2000, 2000] \: \SI{}{\micro\meter}$, sampled at \SI{112.5}{\micro\meter} (i.e., the radius of an Argus II electrode).
This led to a manageable set size ($|\mathcal{T}| \approx 2000$) while still allowing electrodes to be placed right next to each other (if desirable).
To find the electrode $d_i$ at each iteration in Eqs.~\ref{eq:dict-baseline},~\ref{eq:dict-no-overlap}, and \ref{eq:dict-max-dist} above, we thus performed a grid search.

\subsubsection{Stopping Criteria}

The dictionary search was stopped when at least one of the following criteria were met:
\begin{itemize}[topsep=0pt]
    \item visual subfield coverage reached \SI{99}{\percent},
    \item the utility score $F$ did not improve by $\ge \num{1e-6}$ on two consecutive runs,
    \item no more viable electrode locations were available (i.e., $\mathcal{T} \setminus \mathcal{D} \; \mathrm{s.t.} \; ||d, d_i||_2 \ge c \; \forall d_i \: \in \: \mathcal{D} \setminus d = \emptyset$).
\end{itemize}

\section{Results}
\subsection{Visual Subfield Coverage}

\begin{figure}[!t]
    \centering
    \includegraphics[width=\textwidth]{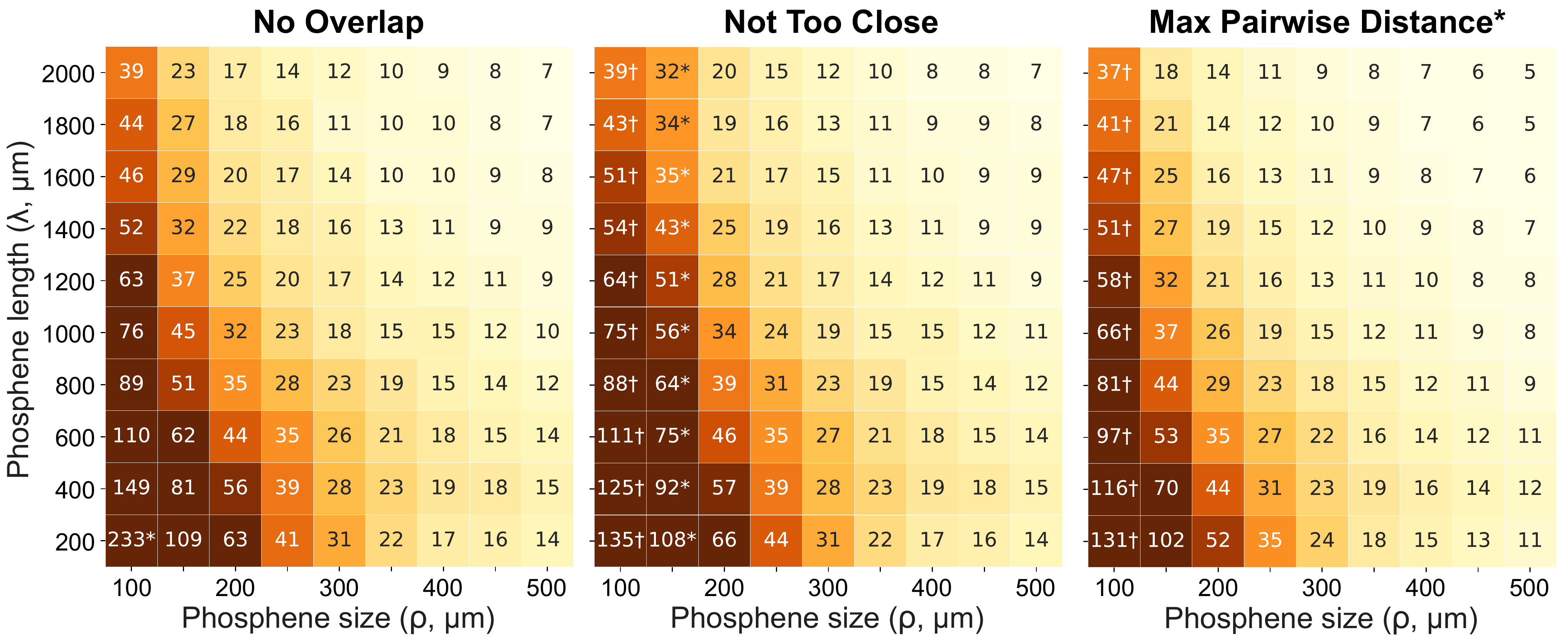}
    \caption {Number of electrodes needed to reach maximum visual subfield coverage for the three dictionary selection strategies ($\dagger$: $80-\SI{95}{\percent}$ coverage, $*$: $95-\SI{99}{\percent}$ coverage).
    Max Pairwise Distance always had $*$, unless otherwise noted by $\dagger$.}
    \label{fig:results-num-electrodes}
\end{figure}

The results of the greedy dictionary selection are shown in Fig.~\ref{fig:results-num-electrodes}.
For all three dictionary selection strategies, the number of electrodes required to cover at least \SI{99}{\percent} of the visual subfield was inversely proportional to $\rho$ and $\lambda$.
As expected, the largest number was achieved with the smallest, most compact phosphene shape ($\rho=\SI{100}{\micro\meter}, \lambda=\SI{200}{\micro\meter}$).
The required electrode number dropped rapidly with increasing $\rho$ and $\lambda$, indicating that for large phosphenes, a prototype implant such as Argus I ($4 \times 4$ electrodes) might be sufficient to cover the whole subfield.

It is interesting to note that, for any given $\rho$ and $\lambda$ combination, the Max Pairwise Distance strategy required a smaller number of electrodes than the No Overlap strategy.
However, using the Not Too Close strategy, smaller $\rho$ and $\lambda$ combinations were no longer able to reach full coverage, as electrodes could no longer be placed too close to each other or to the visual subfield boundary.
This issue was amplified with the Max Pairwise Distance strategy, with which coverage dropped for most $\rho$ and $\lambda$ combinations to \SI{95}{\percent}.

\begin{figure}[!t]
    \centering
    \includegraphics[width=\textwidth]{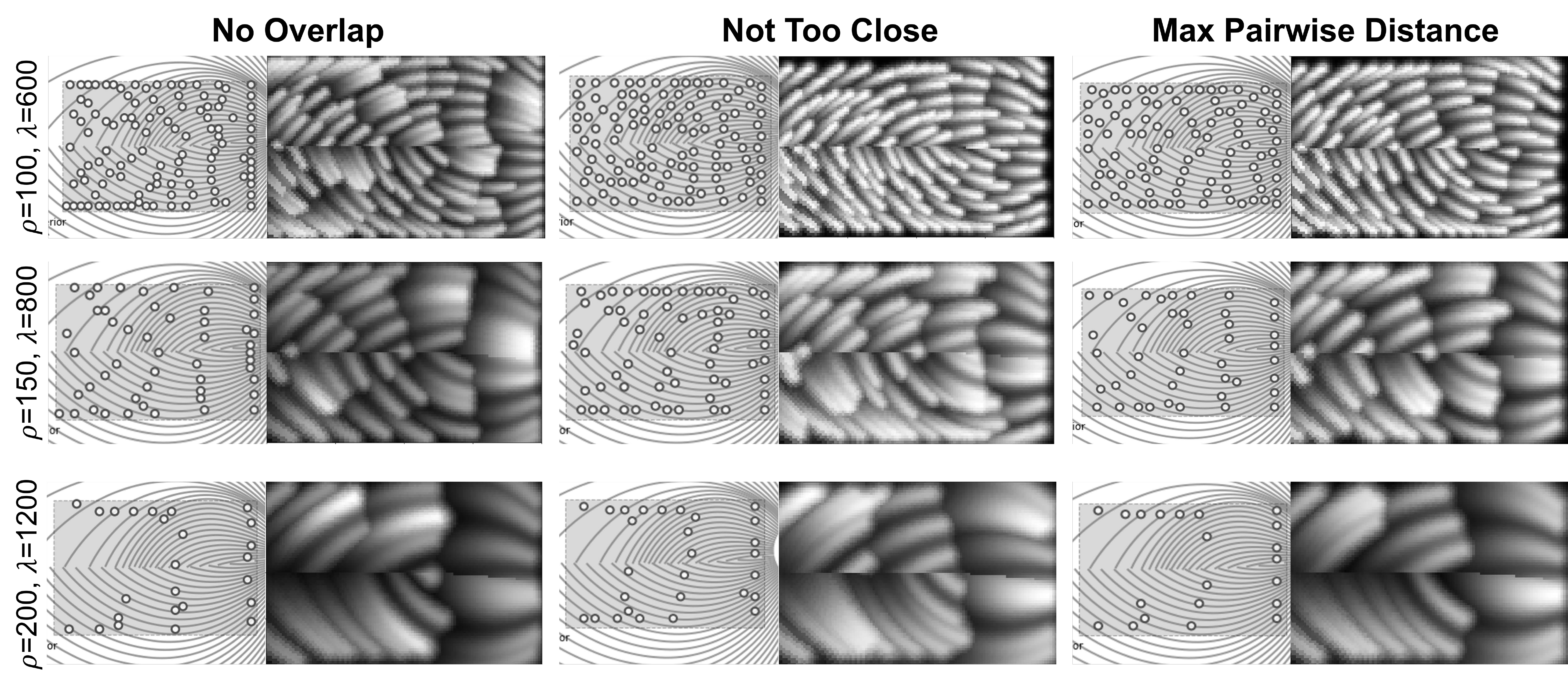}
    \caption{Representative examples of the final electrode arrangements generated with the three dictionary selection strategies for different $\rho$ and $\lambda$ combinations.
    The left half of each panel shows the retinal location of all electrodes (small circles) in the implant, and the shaded region indicates the visual subfield (compare to Fig.~\ref{fig:axonmap}). The visual percept that results from simultaneously activating every electrode in the implant is shown in the right half of each panel.
    }
    \label{fig:results-final-implants}
\end{figure}

With larger $\rho$ and $\lambda$ values, an implant required less than ten electrodes to cover the visual subfield.
However, such a small number was not necessarily desirable, as it also reduced the number of distinct phosphenes that the implant can produce.
Rather than focusing on visual subfield coverage alone, one might therefore ask what kinds of electrode arrangements the three dictionary selection strategies yield and what the resulting percepts look like.

\subsection{Electrode Arrangement}

Example electrode arrangements suggested by the three dictionary selection strategies are shown in Fig.~\ref{fig:results-final-implants}.
Here it is evident that, as $\lambda$ increased, electrodes were preferentially placed on the top, bottom, and right boundaries of the visual subfield.
This placement would often lead to the longest streaks, thus yielding the largest coverage.
For small $\rho$ values, electrodes aggregated mainly on \acp{NFB}, spaced $\lambda$ apart, so that the streaks generated by different electrodes tiled the visual subfield.
As $\rho$ increased, electrodes tended to migrate away from the border and more inward, due to the outward spread of the generated percept.

\begin{figure}[!t]
    \centering
    \includegraphics[width=\linewidth]{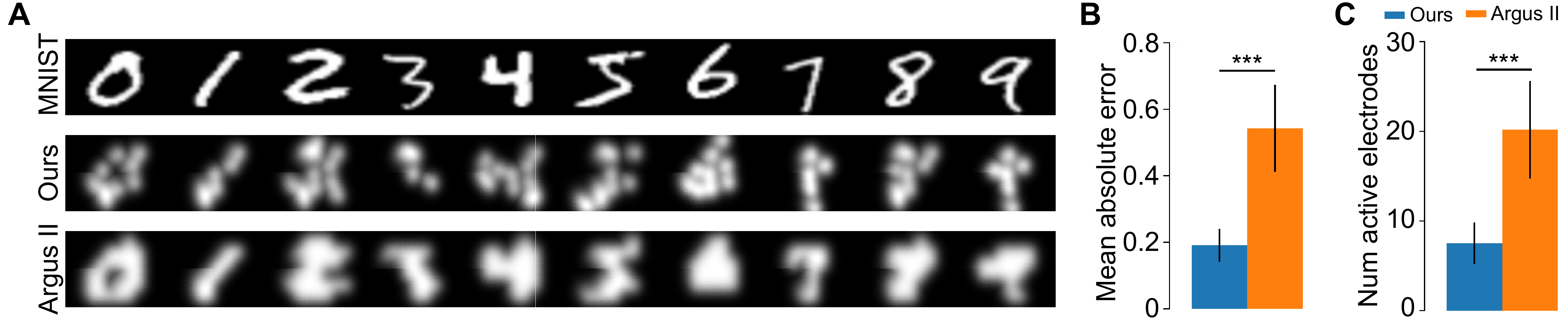}
    \caption{\textbf{A)} Representative samples from the MNIST database (\emph{top row}) represented with an optimized 60-electrode implant (``No Overlap'', \emph{middle row}) and Argus II (\emph{bottom row}).
    \textbf{B)} Mean absolute error for 100 randomly selected MNIST digits (vertical bars: standard error of the mean, ***: $p<.001$).
    \textbf{C)} Number of active electrodes needed to represent the 100 digits.
    All simulations had $\rho=\SI{200}{\micro\meter}, \lambda=\SI{400}{\micro\meter}$.}
    \label{fig:results-mnist}
\end{figure}

\subsection{Comparison with Argus II}

To assess whether the optimized electrode arrangements
provided an improvement over a rectangular \ac{MEA}, we compared our results to Argus II.
We thus modified our experiment such that the dictionary consisted of at most 60 electrodes (i.e., same as Argus II).
We then considered the ability of the optimized implant to represent handwritten digits from the MNIST database (Fig.~\ref{fig:results-mnist}A).
We found that the optimized implant not only led to smaller errors between predicted percepts and ground-truth digits (paired $t$-test, $p<.001$; Fig.~\ref{fig:results-mnist}B) but also required less active electrodes overall (paired $t$-test, $p<.001$; Fig.~\ref{fig:results-mnist}C).

Furthermore, we calculated the visual subfield coverage achieved with these 60 electrodes and compared the result to Argus II (Table.~\ref{tab:Argus II Comparison Results}).
Given the fixed number of electrodes, our electrode selection strategies were always able to cover a larger portion of the visual subfield than Argus II.

Overall these results suggest that a rectangular grid might not be the best electrode arrangement for epiretinal prostheses.

\section{Conclusion}

We report epiretinal electrode arrangements that maximize visual subfield coverage as discovered by dictionary learning. We were able to obtain nearly full coverage across different $\rho$ and $\lambda$ values for a given implant size and electrode radius, and report the number of electrodes necessary to reach this coverage.
Future work should extend the dictionary selection strategy to other implant sizes and explore all possible stimuli, not just the ones that may result in sufficient visual subfield coverage.

This preliminary study is a first step towards the use of computer simulations in the prototyping of novel neuroprostheses.
To the best of our knowledge, this is the first study using a psychophysically validated phosphene model to optimize electrode arrangement.
Even though we have focused on a specific implant technology, our strategy can be readily applied wherever there is a topological mapping from stimulus space to perceptual space (e.g., visual, auditory, tactile).
This means that our approach could be extended to other neuromodulation technologies that include (but are not limited to) other electronic prostheses and optogenetic technologies.
In the near future, our work may therefore guide the prototyping of next-generation epiretinal prostheses.

\begin{table}[!t]
\def\arraystretch{1.2}%
\setlength{\tabcolsep}{6 pt}
\noindent \begin{tabular}{r|rrr|rrr|rrr}
\hline
\multicolumn{1}{r}{$\rho$}  & \multicolumn{3}{c|}{{ \SI{100}{\micro\meter}}} & \multicolumn{3}{c|}{{ \SI{150}{\micro\meter}}} & \multicolumn{3}{c}{{ \SI{200}{\micro\meter}}} \\
\hline
{$\lambda$}  & AII & NTC & MPD & AII & NTC & MPD & AII & NTC & MPD \\
\cline{2-10}
{\SI{200}{\micro\meter}} & 39.2	& 43.2	& 42.9	& 74.1	& 75.6	& 75.9	& 86.2	& $>$99* &	95.2 \\

{\SI{400}{\micro\meter}}	& 55.0	& 66.8 & 66.8	& 80.6	& 91.3	& 91.4	& 87.5	& $>$99*	& 95.1	 \\

{\SI{600}{\micro\meter}} & 62.4 &	81.8 &	81.3 &	81.9 &	96.9 &	95.3 &	88.4 &	$>$99* & 95.2 \\

{\SI{800}{\micro\meter}} & 66.7 &	88.8 &	88.8 &	82.8 &	97.7 &	95.2 &	89.0 &	$>$99* & 95.4 \\

{\SI{1000}{\micro\meter}} & 69.4 &	92.5 &	92.2 &	83.5 &	98.1* &	95.2 &	89.6 &	$>$99* &	95.8 \\

{\SI{1200}{\micro\meter}} & 71.9 &	92.6 &	92.5 &	84.2 &	98.0* &	95.3 &	90.1 &	$>$99* &	95.7 \\

{\SI{1400}{\micro\meter}} & 72.4 &	93.9* &	93.8 &	84.6 &	98.1* &	95.27 &	90.4 &	$>$99* &	95.6 \\

{\SI{1600}{\micro\meter}} & 73.4 &	93.9* &	93.8 &	85.0 &	98.5* &	95.41 &	90.7 &	$>$99* &	95.8 \\

{\SI{1800}{\micro\meter}} & 74.1 &	94.0* &	93.9 &	85.4 &	98.6* &	95.16 &	91.0 &	$>$99* &	95.7  \\

{\SI{2000}{\micro\meter}} & 74.7 &	94.4* &	94.4 &	85.9 &	98.3* &	95.08 &	91.4 &	$>$99* &	95.2 \\
\hline
\end{tabular} \\
\caption{Visual subfield coverage achieved with 60 electrodes using different electrode placement strategies. AII: Argus II. NTC: Not Too Close. MPD: Max Pairwise Distance. Asterisk (*): Search terminated before utilizing 60 electrodes.}
\label{tab:Argus II Comparison Results} 
\end{table}

\section*{Acknowledgements}

This work was supported by the National Institutes of Health (NIH R00 EY-029329 to MB).
The authors would like to thank Madori Spiker and Alex Rasla for valuable contributions to an earlier version of the project.

\bibliographystyle{splncs04}
\bibliography{paper1645}

\end{document}